\documentstyle[12pt,spin98_desy,epsfig]{article}

\def\be{\begin{equation}}
\def\ee{\end{equation}}
\def\bea{\begin{eqnarray}}
\def\eea{\end{eqnarray}}

\begin{document}

\title{ACCELERATION AND STORAGE OF POLARIZED ELECTRON BEAMS
\footnote{Invited plenary talk presented at the 13th International Symposium on
High Energy Spin Physics (SPIN98), Protvino, Russia, September 1998.
Also as DESY report 98--182.}
}

\author{D.P. BARBER}

\address{Deutsches Elektronen--Synchrotron, DESY, \\
 22603 Hamburg, Germany. \\E-mail: mpybar@mail.desy.de}


\maketitle

\abstracts{I review recent developments in spin dynamics in electron 
storage rings and accelerators.}

\section{Introduction}
This article provides an update on my review at SPIN96 \cite{spin96} on
activities surrounding spin polarization
in electron storage rings and accelerators. In the written versions
of previous talks at the Spin Symposia I have opened with a 
review of the basic theory of radiative spin polarization, spin precession
and resonance phenomena.
That background material is readily available in the
proceedings of earlier Symposia and elsewhere  
\cite{spin96,spin94,mont98,hbook}. 
So to avoid repetition I will, on this occasion, launch straight into 
the main themes. Historical overviews of radiative polarization can be
found in \cite{spin96,shat90}. 
\section{High energy storage rings: HERA and LEP}
HERA is the $e^{\pm}-p$ collider at DESY in Hamburg. The $e^{+}$ or $e^{-}$
beams run at about 27.5 GeV. Up to the end of 1997 the proton ring
ran at 820 GeV. In 1998 it has been running at 920 GeV. 
$e^{\pm}$ beams in storage rings can become  vertically polarized  by the  
Sokolov--Ternov effect (ST) \cite{st64,mont98} and a
key aspect of HERA is that since 1994 longitudinal spin polarization has been
supplied to the HERMES experiment \cite{bruell} with the help of a pair of 
spin rotators \cite{bar95}. 

The {\it value} of the polarization in an  $e^{\pm}$ storage ring is
the same everywhere around the ring even with rotators running. However, 
at high energy, as at HERA, the polarization is  very 
sensitive to the size and form of closed orbit distortions. With very 
careful adjustment of the vertical closed orbit distortion using
{\it harmonic closed orbit spin matching} \cite{hbook}, up to about 70 \% 
polarization has been seen at HERA with the HERMES rotators running. This is
to be compared  with the theoretical maximum for that configuration of 
89.06 \% . 

The polarization at HERA can also be affected by the beam--beam (b--b) forces 
due to 
collisions with the  proton beam at the H1 and ZEUS experiments where, 
incidently, the polarization is vertical. Since 
the b--b forces are very nonlinear it is very difficult to make
analytical calculations of their  effects on $e^{\pm}$ beams.
And of course, it is even more difficult to make analytical estimates of the 
effects on the polarization. However, the naive 
expectation is that the b--b forces reduce the polarization and some
spin--orbit tracking calculations support that view \cite{blpac95}. 
Normally is it assumed that it is a good idea to reduce the b-b tune shift
(explained  below) but as usual, there is no substitute for measurement and  
in 1996 even during collisions with 50 mA of protons, positron 
polarizations of about 70 \% were observed with the rotators running. 
One such run lasted ten hours.
So, at least in {\it those} optics, b--b forces had little influence.

Since a few proton bunches, which would normally be in collision with 
electrons  (positrons), are by intent missing, not all electron (positron)
bunches come to collision with protons. Towards the end of 1996 a second 
polarimeter, built by HERMES, came into operation \cite{mostspin96}. 
In contrast to the original
polarimeter which measures the level of vertical polarization in the West 
area by Compton scattering using the so called single photon technique
\cite{nim94},  the  new polarimeter, which employs Compton scattering to
measure longitudinal polarization directly close to HERMES and which uses the
multi--photon technique, can collect data more quickly. It then became 
possible to study the positron polarization with sufficient precision
on a bunch--to--bunch basis. Figure 1 summarizes a typical measurement 
for collisions of positrons with  about 60 mA of protons
\begin{figure}[h]
\begin{center}
  \setlength{\unitlength}{1mm}
\epsfig{file=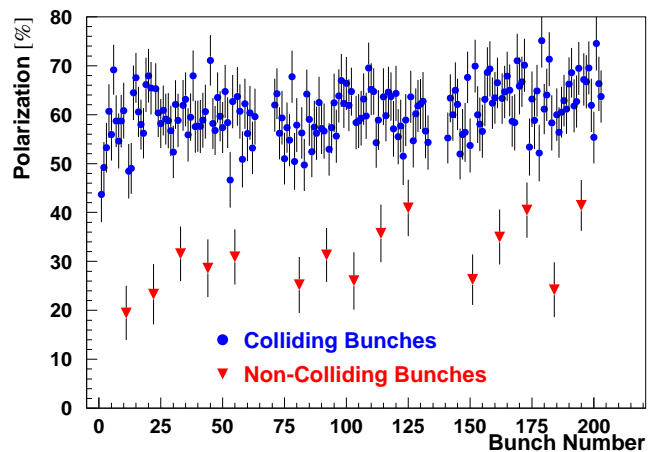,height=7.1cm}
\end{center}
\caption{An example of a measurement with the longitudinal polarimeter
of HERMES of the polarization of positrons 
colliding/not colliding with protons at HERA.}
\label{fg:nuplim}
\end{figure}
and in this example, contrary to intuition,  the colliding bunches have
{\it more} polarization than the non-colliding bunches. At present we 
interpret this unexpected result as being due to the b--b tune shift: an 
oncoming proton bunch appears to the positrons as a nonlinear lens and
to a first approximation the colliding positron bunches have betatron tunes 
which differ from those of the non--colliding bunches. 
So, by the routine adjustment of some quadrupole strengths to get overall
betatron tunes which lead to  optically stable  running conditions for the
colliding bunches and to high polarization (averaged over the bunches),
it is possible that the non--colliding bunches
are close to a  depolarizing spin--orbit resonance (probably a synchrotron
sideband resonance of a parent resonance \cite{hbook})
and likely that the colliding bunches are not on such a resonance.
This interpretation is supported by the fact that  
on other occasions with slightly  different machine tunes, there is either 
little difference between the colliding and non--colliding polarizations or
the colliding bunches indeed have less 
polarization than the non--colliding bunches. For the measurement of figure 1  
the vertical b--b tune shift was about 0.034 for each interaction point.

Apart from the sensitivity to orbital tunes one observes that in the  
presence of b--b effect the rise time for the polarization after injection is
sometimes larger than that expected from standard radiative polarization 
theory and that the polarization level is sometimes relatively insensitive to
the settings of the closed orbit 
harmonics of the harmonic closed orbit correction scheme \cite{eg97}. 
Naturally, since the b--b effect can affect the rise time
it makes little sense to calibrate a polarimeter by measuring the
rise time after resonant depolarization \cite{nim94} while the 
beam is in collision with protons.

The electron (positron) bunches in HERA come in three groups of about sixty
bunches with 
gaps between the groups. This causes dynamic beam loading of the rf cavity
system needed to replace the energy lost by radiation. That in turn can 
cause the synchrotron tune to vary along a group with the result that
electrons (positrons) at the beginning of a group can be closer to a
depolarizing 
resonance than those at the end (or vice versa). Thus we sometimes see a 
variation  of the
polarization of the colliding bunches across a group.

        In 1997 under normal running conditions with typically 80 mA of
protons we had about 50 \%  polarization, averaged over the bunches. 
Even towards the end of the year when we ran with over 100 mA of protons
(vertical b--b tune shift $\approx 0.035$) a polarization level of 50 \% could 
still be reached.
It might have been possible to attain more with careful
adjustment of the closed orbit  but we must normally make a compromise between
tuning the orbit and providing stable running conditions for the high energy 
physics experiments.  

Electrons and positrons can also become  polarized in LEP, the $e^{\pm}$  
collider at CERN in Geneva. 
The effect of b--b forces on polarization
has also been studied there and it has been found that the polarization is 
very sensitive to optical parameters \cite{assspin94}, just as at HERA. 

So far, we cannot claim that we understand in detail all the effects of 
b--b forces on the polarization and it has not yet been conclusively 
demonstrated that it is impossible to get high polarizations in the
presence of a b--b effect which is large but not large enough to disrupt the
beam itself.
More investigations under controlled and reproducible conditions are needed.

        In the winter shutdown 1999/2000 we plan to change the
geometry of the North and South interaction regions of HERA in order to
 increase the luminosity supplied to the H1 and ZEUS experiments by 
a factor of about 4.7 beyond the 
design value of  1.5$\cdot 10^{31}$ cm$^{-2}$sec$^{-1}$ \cite{egepac98}. 
This will be achieved by reducing the beam cross--sections at the interaction
points (IP's) and by reaching the design currents. The smaller beam sizes will
be achieved by having smaller $\beta$ 
functions at the IP's and by changing the optics in the arcs in order to 
decrease the horizontal emittance. 
These changes have profound consequences for $e^{\pm}$
polarization: smaller $\beta$ functions imply that the focussing magnets must
be moved closer to the IP's and this in turn means that  
the ``antisolenoids'' which currently compensate the H1 and ZEUS
experimental solenoids will be removed. In fact  new  stronger combined
quadrupole and dipole magnets will be installed on each side of the H1 and
ZEUS IP's and their fields will overlap with the solenoid fields. 
At the same time additional spin rotators will be installed to enable H1 
and ZEUS to run with longitudinal polarization. We  plan to run these 
rotators in a slightly mistuned state designed so that they effectively 
compensate for the effect on the equilibrium spin axis 
of the overlap of solenoid and dipole fields and ensure that the spin axis is 
still vertical in the arcs --- an essential requirement for high 
polarization \cite{hbook}. The absence of the antisolenoids means that the
 resultant
orbital coupling must be corrected away with skew quadrupoles and that the
computer programs involved in the 
{\it strong spin matching} \cite{hbook} and calculation of polarization must 
be upgraded to handle the new and complicated magnetic field configurations
near the IP's. Accounts of the full implications for the maintenance
of radiative polarization  can be found in \cite{egepac98,mbepac98}.

Since the ratio:
(depolarization rate/polarization rate) \cite{hbook} rises strongly with 
energy,
it was much more difficult to attain high polarization in LEP at the old
running energy of about 46 GeV per beam (near the $Z^0$) than at HERA with
27.5 GeV. Moreover the vertical polarization of LEP (there are no spin 
rotators) is of little use for high energy physics and in any case the
rise time for the polarization is a few hours compared with the twenty 
minutes of HERA. In spite of these difficulties the LEP team recorded a
polarization of about 57 \% in 1993 \cite{assspin94,bdspin96} --- a major
achievement. Under routine running conditions 
at about 46 GeV, LEP ran with 5 -- 10 \% polarization.
But this was sufficient for the exploitation of polarization to measure the
beam energies, and hence the $Z^0$ mass,
by means of  resonant depolarization,  leading to a precision of about
1.5 MeV \cite{zfp95,bdspin96}.

But now LEP runs at above 80 GeV per beam and the polarization is effectively
zero; 5 \% polarization was recorded at 55.3 GeV but this was down to 2 \%
at 60.6 GeV. However, vertical polarization can still be used for energy
calibration --- but indirectly by calibrating a flux loop and  sixteen NMR 
probes in  dipoles at about 41, 45, 50 and 55 GeV \cite{mpepac98}
and then using the calibrated flux loop and NMR probes at above 80 GeV. 
The estimated systematic error for this method is about
25 MeV per beam but the long term aim is for a precision of about 15 MeV per
beam. 
\section{Accelerators}
Because the rise time for ST polarization is typically in the range of 
minutes to many hours, extracted polarized $e^{-}$ beams can only be obtained
from accelerators by injecting a pre-polarized beam from 
a source. 
Modern gallium--arsenide sources \cite{mulspin96} deliver up to about  
80 \% electron polarization. But that must then be preserved during
acceleration. There are at present no suitable polarized positron sources
and therefore no  extracted polarized positron beams. 

When dealing with polarized beams we can distinguish two basic types of  
accelerators, namely linear accelerators where, by design, the particle
velocity and the accelerating electric field are essentially parallel, and 
ring accelerators where, in addition, the beam must make many thousands  
or even millions of turns in the magnetic guide field on the way to full
energy.  
If the particle velocity and the electric field are almost parallel, then 
according to the T--BMT precession equation \cite{hbook, mont98} there is
very little spin precession and hence little opportunity for depolarization.
The (lack of) spin precession in the two mile long accelerating section of 
the SLC at SLAC
in California is the prime example of this.
The SLC has regularly delivered
an electron beam of about 46 GeV with over 70 \% polarization \cite{mulspin96}.

A good example of the other type is ELSA \cite{elsaepac98}, the 3.5 GeV ring
at Bonn, Germany which accelerates vertically polarized electrons.
According to the T--BMT  equation, in vertical  
magnetic fields, spins precess
$a \gamma$ times per turn where $a = (g-2)/2$ is the gyromagnetic anomaly 
and $\gamma$ is the Lorentz factor.
If the spin precession  is in resonance with the orbital motion:
$a \gamma = m_0 + m_x\,Q_x + m_z\,Q_z + m_s\,Q_s$ where the $m$'s are integers 
and the $Q$'s are orbital tunes,
the spins can be strongly disturbed and the polarization can be 
lost. 
Since the precession rate $a \gamma$ is proportional to the energy, and
increases by unity for every 440 MeV increase in energy, several such 
resonances must be crossed on the way to 3.5 GeV. A typical example  
is at 1.32 GeV in ELSA. This corresponds to $m_0 = 3$ but $m_x = m_z = m_s =0$.
Spin perturbations in this case result from the radial fields ``seen'' by
the spins in the quadrupoles when there is vertical closed orbit distortion.
A first approximation for the polarization surviving the crossing of a 
resonance is given by the Froissart--Stora (FS) 
formula \cite{fs,roserspin94}:
\begin{eqnarray} 
\frac {P_{\rm final}}{P_{\rm initial}}  =
                             2~e^{-\frac{\pi {{| \epsilon |} ^2}}{2 \alpha}} -1
\end{eqnarray}
where $\epsilon$ is the ``resonance strength'', a measure of the dominant 
spin perturbation at resonance, and $\alpha$ expresses the rate of 
resonance crossing. Thus if the resonance is crossed sufficiently  quickly
(${{| \epsilon |} ^2}/{ \alpha}$ is small) the
polarization is hardly affected but if it is crossed sufficiently slowly
(${{| \epsilon |} ^2}/{ \alpha}$ is large) a complete reversal of the vertical
polarization can occur without much change in the magnitude.
Measurements of the surviving polarization for  a range of
${{| \epsilon |}^2}/{ \alpha}$ values are now available from ELSA 
\cite{elsaepac98} both for 1.32 GeV and for 1.76 GeV ($m_0 = 4$).
The measurements for $m_0 = 3$ show good agreement with the prediction
of the FS formula. In particular, by running at
${{| \epsilon |}^2}/{ \alpha} \ge 4.0 $ one can preserve the value of the
polarization by
means of complete spin flip. However, for $m_0 = 4$ only partial spin flip 
with a $|P_{\rm final}/P_{\rm initial}| \approx 0.8$  is seen even out to
${{| \epsilon |}^2}/{ \alpha} \approx 12.0$. This is probably
due to the encroachment of stochastic spin decoherence  
owing to synchrotron radiation emission at the higher energy. If this is indeed
the case, these measurements provide a window on what can be expected from
attempts to flip $e^{\pm}$ spins in HERA \cite{decohspin94,khdecoh97}. 

A good compromise between the space occupied by the SLC and the spin 
perturbation problems of ELSA, is provided by the ring at the Jefferson 
Laboratory in Virginia, U.S.A. This was designed to provide longitudinally
polarized electrons at 4 GeV.
However, it is already providing up to 77 \% polarization at 
5 GeV \cite{ruttpriceprivcom}. This ring 
combines the best of both worlds; it consists essentially
of two parallel superconducting linear accelerators connected at their 
ends by semicircular arcs of bending magnets. The beam is accelerated to 
full energy in just five turns. In the arcs the energy is constant so that
there is no resonance crossing and in the accelerating sections, just as in 
the SLC, spin perturbations are negligible. In any case, with so few turns and 
with such a large acceleration rate (large $\alpha$) no depolarization 
is expected.  This machine is already a wonderful tool for research with
spin and in the long term with steady improvements it might even be possible 
to reach  12 GeV \cite{ruttpriceprivcom}.
\section{Kinetic polarization} 
At SPIN96 \cite{spin96} I reported on progress towards obtaining longitudinal
electron
polarization in the AmPs ring in Amsterdam \cite{shat96}. This ring runs at up
to 900 MeV. The electron beam is injected pre--polarized and a Siberian Snake,
based
on a superconducting solenoid, is employed to  stabilize the polarization and 
to ensure that the polarization is longitudinal at the internal target.
A fascinating and educational aspect of this machine is that, because the 
normal radiative polarization process is eliminated owing to the fact
that the equilibrium polarization lies in the horizontal plane,  
a weaker polarization mechanism, ``kinetic polarization'', might become
observable \cite{mont84}. As reported at this Symposium by 
Yu. Shatunov, measurements at AmPs have now provided preliminary 
experimental evidence for this effect  \cite{shat98}. Confirmation of
these  observations will vindicate efforts \cite{mont98,dk73,mane87} to put 
the theory of the combined radiative polarization 
and radiative depolarization processes on a firm semiclassical basis. 
Moreover, kinetic polarization is expected to contribute $+/-$
a few percent to the $e^{\pm}$ polarization at HERA when the 
spin rotators are running. But since its magnitude depends sensitively
on the details of the closed orbit distortion and since that cannot be
measured with sufficient accuracy, kinetic polarization sets a limit to the
precision with which the polarimeters can be calibrated by measuring the
polarization rise time \cite{nim94,bar95} with the rotators in use.

Perhaps the work at AmPs can be extended at the B$\tau$CF ring being planned in
Beijing \cite{wangdong} and at the MIT--Bates ring \cite{zwart}.

\section{``Spinlight''}
In high energy storage rings, $e^{\pm}$ polarization is normally measured 
using Compton scattering. In linear accelerators Moeller \cite{moeller}
scattering, which is destructive, can be used too. However, there is another
possibility, namely to measure the tiny, O($\hbar$), component of
synchrotron radiation (``spin light'') which depends on 
the orientation of the spins \cite{novosib98}. This causes a small
difference between the spectra of very high energy photons radiated from 
vertically polarized and  unpolarized bunches.
The difference, which has already been detected at low energy
\cite{novosib84}, is proportional
to the polarization and therefore supplies a way of measuring the latter.
Indeed, a feasibility study for a ``spin light polarimeter'' at HERA 
is now being undertaken by physicists from the Yerevan Physics Institute in
Armenia and from DESY \cite{aaspin96}.
Furthermore, the correlation between the radiation spectra and the spin 
orientation 
lies at the heart of the kinetic polarization effect \cite{mont84}
so that even apart from the question of polarimetry, it would be of 
interest to make more detailed measurements and at high energy.

\section{Fokker--Planck theory for spin diffusion}
Now, to conclude, I would like to mention that a way has recently been found, 
using classical concepts,
to write a diffusion equation describing stochastic spin dynamics in storage 
rings. The key is to work with the density in phase space of the spin 
angular momentum as a parallel to the use of particle density for orbital
diffusion. If the Fokker--Planck equation for the orbital motion is known,
the corresponding equation for spin can be written immediately.
More details can be found in another article in these 
proceedings \cite{barheispin98}.

\section*{Conclusion}
$e^{\pm}$ polarization in storage rings and accelerators is an active, 
developing and exciting field. Much is now routine but there are still many
aspects to investigate and challenges to meet.
\section*{Acknowledgments}
I would like to thank E. Gianfelice--Wendt for her valuable comments on HERA
performance; colleagues from the HERA Polarimeter Group for supplying me with
information; 
P. Rutt and J.S. Price for updating me on the
Jefferson Laboratory machine; and  C. Prescott and Yu. Shatunov for 
providing me with information about their  work.
Finally I thank M. Berglund and  M. Vogt for their careful reading of the
manuscript.

\end{document}